\documentclass[12pt,preprint]{aastex}

\begin{document}

\title{The First Measurement of Cassiopeia A's Forward Shock Expansion Rate}

\author{Tracey DeLaney and Lawrence Rudnick}
\affil{Department of Astronomy, University of Minnesota, 116 Church Street 
SE, Minneapolis, MN  55455}
\email{tdelaney@astro.umn.edu, larry@astro.umn.edu}

\begin{abstract}

We have obtained a second epoch observation of the Cassiopeia A supernova 
remnant (SNR) with the \emph{Chandra X-ray Observatory} to measure detailed 
X-ray proper motions for the first time.  Both epoch observations are 50 ks 
exposures of the ACIS-S3 chip and they are separated by 2 years.  
Measurements of the thin X-ray continuum dominated filaments located around 
the edge of the remnant (that are identified with the forward shock) show 
expansion rates from 0.02\% yr$^{-1}$ to 0.33\% yr$^{-1}$.  Many of these 
filaments are therefore significantly decelerated.  Their median value of 
0.21\% yr$^{-1}$ is equal to the median expansion of the bright ring 
(0.21\% yr$^{-1}$) as measured with Einstein and ROSAT.  This presents a 
conundrum if the motion of the bright ring is indicative of the reverse 
shock speed.  We have also re-evaluated the motion of the radio bright ring 
with emphasis on angle-averaged emissivity profiles.  Our new measurement of 
the expansion of the angle-averaged radio bright ring is 0.07 $\pm$ 
0.03\% yr$^{-1}$, somewhat slower than the previous radio measurements 
of 0.11\% yr$^{-1}$ which were sensitive to the motions of small-scale 
features.  We propose that the expansion of the small-scale bright ring 
features in the optical, X-ray, and radio do not represent the expansion of 
the reverse shock, but rather represent a brightness-weighted average of 
ejecta passing through and being decelerated by the reverse shock.  The 
motion of the reverse shock, itself, is then represented by the motion of 
the angle-averaged emissivity profile of the radio bright ring.

\end{abstract}

\keywords{ISM: supernova remnants --- ISM: individual (Cassiopeia A)}

\section{Introduction}
\label{sec:int}

Cassiopeia A (Cas A) is the youngest SNR in the Galaxy and is located 3.4 
kpc away \citep{rhf95}.  The exact age of Cas A is unknown, but recent 
measurements of freely expanding optical knots that lie in front of the 
forward shock give an age of 331 years in \textsc{a.d.} 2002 \citep{tfv01}.  
Proper motion measurements made in the optical, radio, and X-rays show 
variations in expansion rate from 0.1 to 0.3\% yr$^{-1}$ ($v \approx$ 10$^4$ 
km s$^{-1}$) (Thorstensen, Fesen, \& van den Bergh 2001 and references 
therein).  Radio proper motion measurements also show inward motions 
\citep{ar95}.

The forward shock in Cas A has recently been identified in the X-rays as 
a thin, fragmented ring bordering the outer edge of the remnant and 
exhibiting continuum dominated spectra \citep{gkr01}.  The bright ring, 
at 2/3 the radius of the forward shock, is strong in line emission and is 
often associated with the reverse shock.  There is a paradoxical difference 
in expansion rates for the bright ring measured in the different wavebands 
of approximately 3:2:1 for the optical, X-ray, and radio respectively 
(Thorstensen, Fesen, \& van den Bergh 2001 and references therein).

In this paper, we present the first measurement of the proper motions of the 
forward shock fragments in X-rays.  We also present a re-evaluation of the 
motion of the radio bright ring in order to provide a coherent dynamical 
picture.

\section{X-ray Observations, Analysis, and Results}
\label{sec:xobs}

\begin{figure}
\plotone{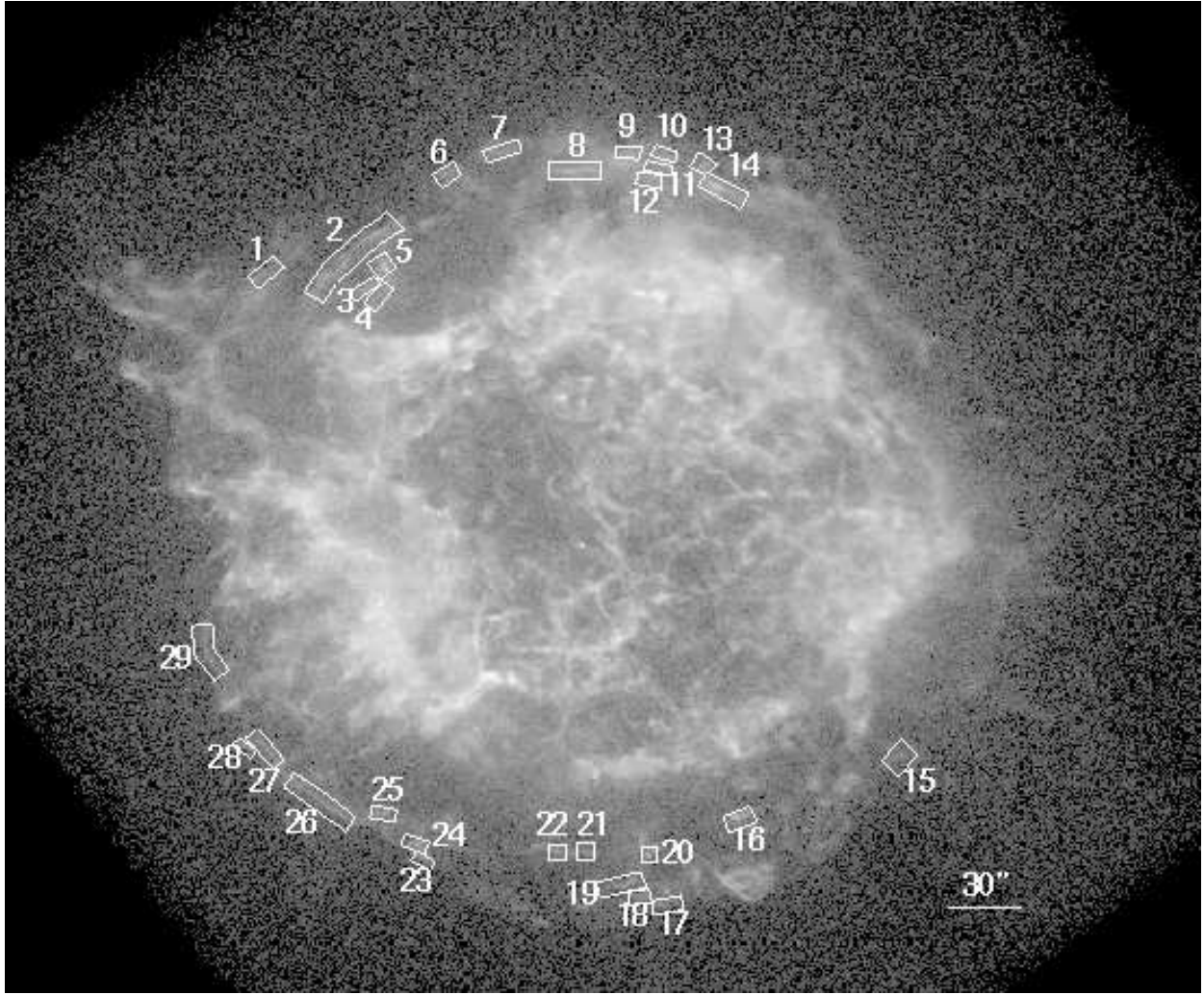}
\caption{The second epoch \emph{Chandra} X-ray image of Cas A (log brightness 
scale) with measured proper motion regions marked.  \label{epoch2}}
\end{figure}

\begin{figure}
\plotone{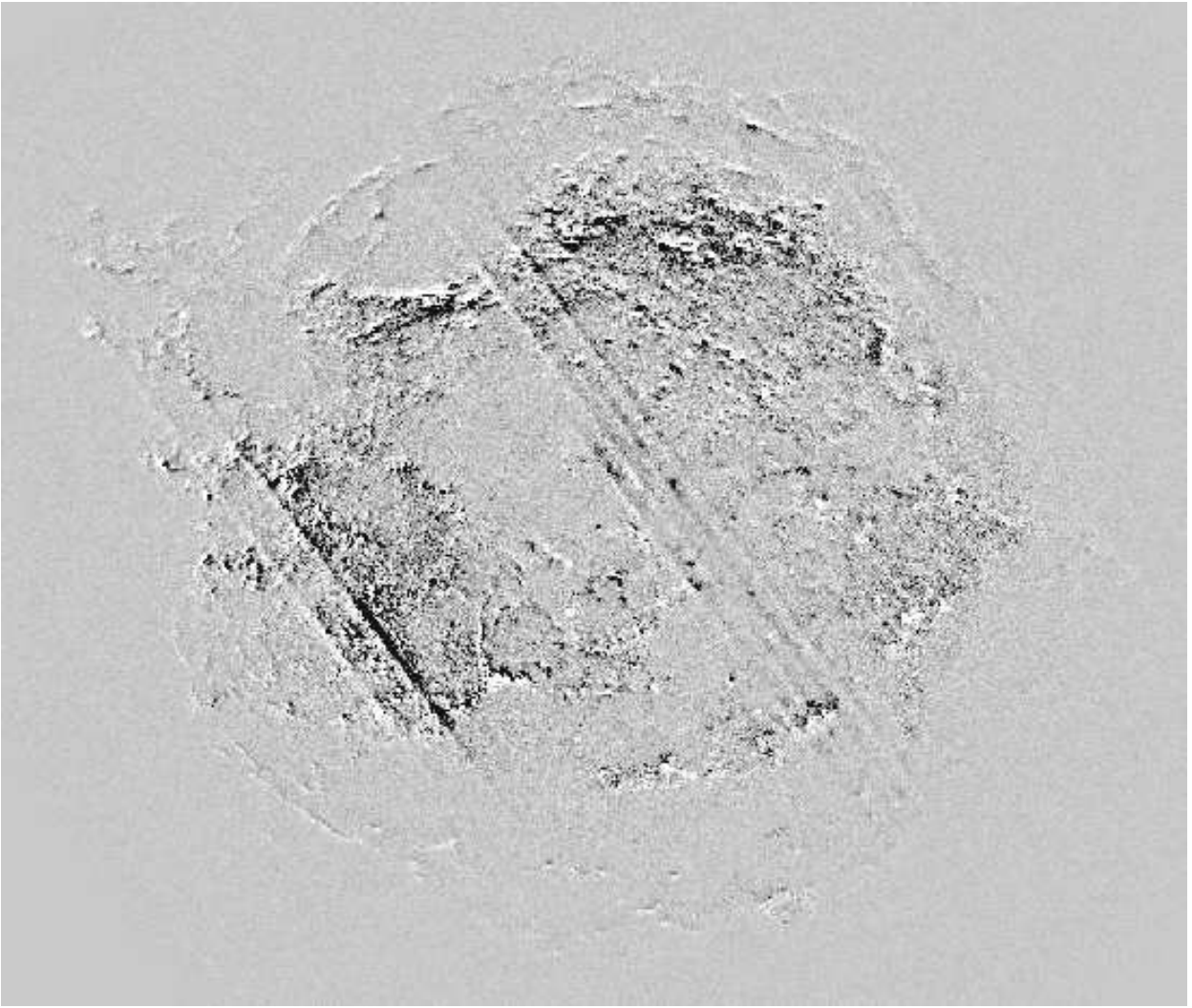}
\caption{Difference image between the second epoch and first epoch 
\emph{Chandra} X-ray images of Cas A.  Bright indicates the 
direction of motion.  The diagonal stripes are due to the node boundaries and 
dead columns on the ACIS-S3 chip. \label{casdiff}}
\end{figure}

We observed Cas A for 50 ks with the ACIS-S3 chip onboard the \emph{Chandra 
X-ray Observatory} on 2002 Feb 6.  The observation parameters are listed in 
Table \ref{tbl1} and were chosen to match those of the first 50 ks exposure 
of the ACIS-S3 chip on 2000 Jan 30-31 \citep{hhp00,gkr01}.  We had to 
recalibrate the first epoch X-ray observation to make use of the newest 
calibration solutions (CALDB version 
2.11)\footnote{\url{http://asc.harvard.edu/ciao/threads/acisgainmap/}} 
and to apply the new geometry solutions (geometry file telD1999-07-23geomN0004.fits)\footnote{\url{http://asc.harvard.edu/ciao/threads/geom\_par/}}.  
We used an energy range of 0.3-10 keV to make the images.  No exposure 
correction has been applied to the images.  Figure \ref{epoch2} is the 
second epoch \emph{Chandra} X-ray image.  The difference between the two 
epochs, after normalization and registration, is shown in Figure 
\ref{casdiff}.  The diagonal stripes on Figure \ref{casdiff} are due to 
the node boundaries and dead columns on the ACIS-S3 chip.

There are two important reasons for requiring the same observational 
parameters, in particular the pointing center and roll angle, for the two 
X-ray observations.  First, the point spread function is asymmetric and a 
function of position on the chip.  This effect would have been very 
difficult, if not impossible, to remove at the accuracy needed for our proper 
motion measurements.  Second, the relative scale in the x and y directions on 
the chip is also uncertain, at a level of 0.07\%.  This could easily 
masquerade as a small but systematic asymmetric expansion.  

A difference in plate scale between the two \emph{Chandra} X-ray observations 
could also affect our proper motion measurements.  In principle, the plate 
scale may change for two reasons -- a change in the mirror focal length and 
a change in the physical chip size. There is no reason to expect a change in 
the physical size of the chip because the temperature was the same for both 
observations.  If the focal length changes by a significant amount, it would 
first result in a deterioration of the point spread function, which is not 
observed (M. Markevitch private communication).

We registered the two images by aligning them on the point source.  The 
second epoch image was first normalized to match the total counts on the 
first epoch image.  The first epoch image was then shifted in RA and DEC 
with respect to the second epoch image to minimize $\chi^2$  of the 
difference between the two images for a 7$\arcsec$ region around the 
point source.  The 68\% confidence limit of the $\chi^2$ distribution leads 
to an uncertainty of 0$\farcs$015 in the registration.  This 
uncertainty translates to 0.005\% yr$^{-1}$ at the mean radius of the 
forward shock fragments (149$\farcs$2).  Although the point source was 
chosen to align the images, it is in motion almost due south at a rate of 
0$\farcs$02 yr$^{-1}$ \citep{tfv01} which results in a maximum error of 
0.013\% yr$^{-1}$ at the mean radius of the forward shock fragments.

We identified the forward shock fragments as tangentially oriented 
filamentary features at or near the outside edge of the remnant with very 
little or no line emission.  However, lack of line emission alone may not be 
an indicator of a forward shock filament as there are interior knots that 
exhibit little line emission as well \citep{hrb00}.  Therefore, our sample 
may have some small contamination from non-forward shock filaments.  In order 
to measure the radial motions of the forward shock fragments, we converted 
each X-ray epoch image to a polar coordinate image using a center 
10$\farcs$12 east and 26$\farcs$33 north of the point source.  This center 
was chosen instead of the expansion center to make the forward shock 
fragments as ``straight'' as possible in polar coordinates so that 
measurement errors due to complicated structures or transverse motions would 
be minimized.  The first epoch image was shifted in radius with respect to 
the second epoch image and $\chi^2$ was computed from the difference between 
the two epoch images at each shift position for each small region outlined 
in Figure \ref{epoch2}.  In order to determine the errors, we made ten Monte 
Carlo simulations of the Poisson noise for each shock fragment and calculated 
the rms scatter of the minimum $\chi^2$ position.  To calculate the 
expansion rate for each region, we used the distance from the optical 
expansion center at $23^{\mathrm{h}}23^{\mathrm{m}}27\fs77$, 
$+58\degr48\arcmin49\farcs4$ (J2000) \citep{tfv01}.  We avoided those 
regions where the node boundary and bad column artifacts would interfere 
with the measurements.  We also avoided radially oriented filaments because 
our measurement method is less sensitive to the motions of such filaments.  As 
a result of these limitations, we have not evenly sampled all azimuth ranges.

\begin{figure}
\plotone{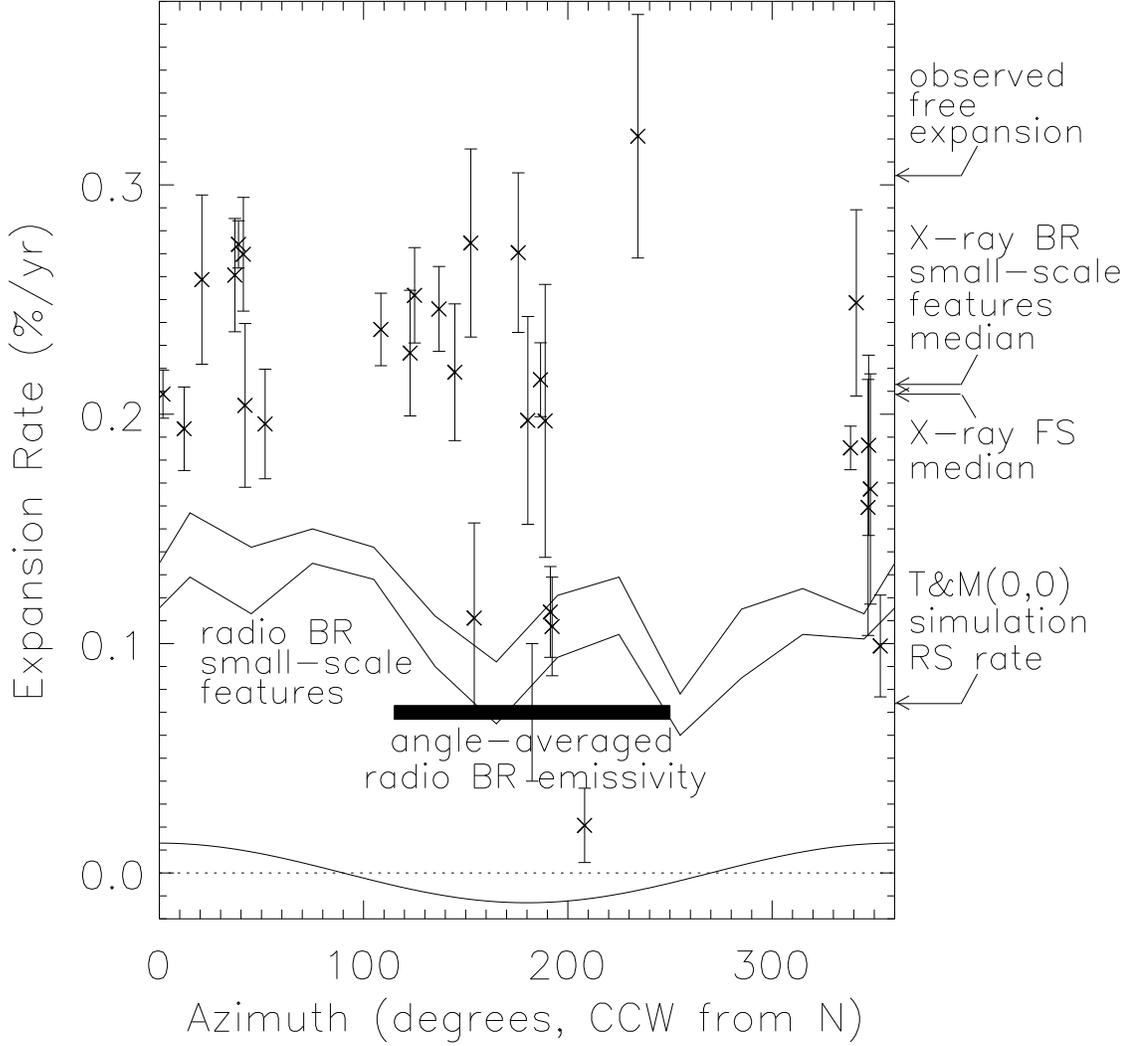}
\caption{Expansion rate vs. azimuth.  Crosses:  forward shock (FS) 
fragments.  Double solid line:  \citet{krg98} radio bright ring (BR) 
small-scale features, line separation shows measurement errors.  Solid 
black rectangle:  angle-averaged radio BR emissivity, length shows azimuth 
range.  Also shown:  observed free expansion rate \citep{tfv01}, median 
expansion rate of the X-ray BR small-scale features \citep{krg98,vbk98}, 
and T\&M(0,0) reverse shock (RS) expansion rate rate \citep{tm99} as 
described in the text.  Sine wave at bottom:  correction to be subtracted 
from the measurement of FS fragments due to the motion of the point source as 
described in the text.  CCW=counterclockwise.  
\label{expplot}}
\end{figure}

Figure \ref{expplot} is a plot of expansion rate \emph{vs.} azimuth for the 
forward shock fragments identified in Figure \ref{epoch2} and listed in 
Table \ref{tbl2}.  The expansion rate of the X-ray forward shock fragments 
ranges from 0.02 to 0.33\% yr$^{-1}$ with a median expansion rate of 
0.21\% yr$^{-1}$.  This is equivalent to a median velocity of the forward 
shock fragments of 5029 km s$^{-1}$ assuming a distance of 3.4 kpc to Cas A 
\citep{rhf95} and no projection effects.  Based on an explosion date of 
\textsc{a.d.} 1671.3 \citep{tfv01}, the median expansion parameter, m, is 
0.69, where $R \propto t^m$ and $m$=1 corresponds to free expansion.  For 
reference, the free expansion rate is indicated in Figure \ref{expplot}.  The 
sine wave at the bottom of Figure \ref{expplot} indicates the expected 
corrections that should be subtracted from our measurements to correct for 
the point source motion as discussed above.

\section{Radio Observations, Analysis, and Results}
\label{sec:robs}

The radio data were taken at the Very Large Array\footnote{The VLA is 
operated by the National Radio Astronomy Observatory, which is a facility of 
the National Science Foundation, operated under cooperative agreement by 
Associated Universities, Inc.}.  The center date of the first epoch radio 
image is 1985.67 and the center date of the second epoch radio image is 
1994.87.  The radio data were taken in B, C, and D configuration at 6 cm.  
For details on the reduction of the radio data, see, for instance, 
\citet{arl91}.

We measured the motion of the bright ring using the angle-averaged brightness 
and emissivity radial profiles.  Angle-averaging enhances the contributions 
of the large-scale and diffuse brightness features relative to the small-scale 
brightness features.  In this manner, the measurements are sensitive to 
changes in the average position of the fine scale ensemble rather than to the 
average of the individual changes.  We have chosen to use the radio images 
instead of the X-ray images to measure motions of the angle-averaged bright 
ring for two reasons.  First, the longer time span between radio observations 
allows us to calculate the expansion rate more accurately.  Second, unlike the 
X-ray bright ring, the southern portion of the radio bright ring is rather 
uniform over a large range in azimuth (115$\degr$ to 250$\degr$).

\begin{figure}
\epsscale{0.7}
\plotone{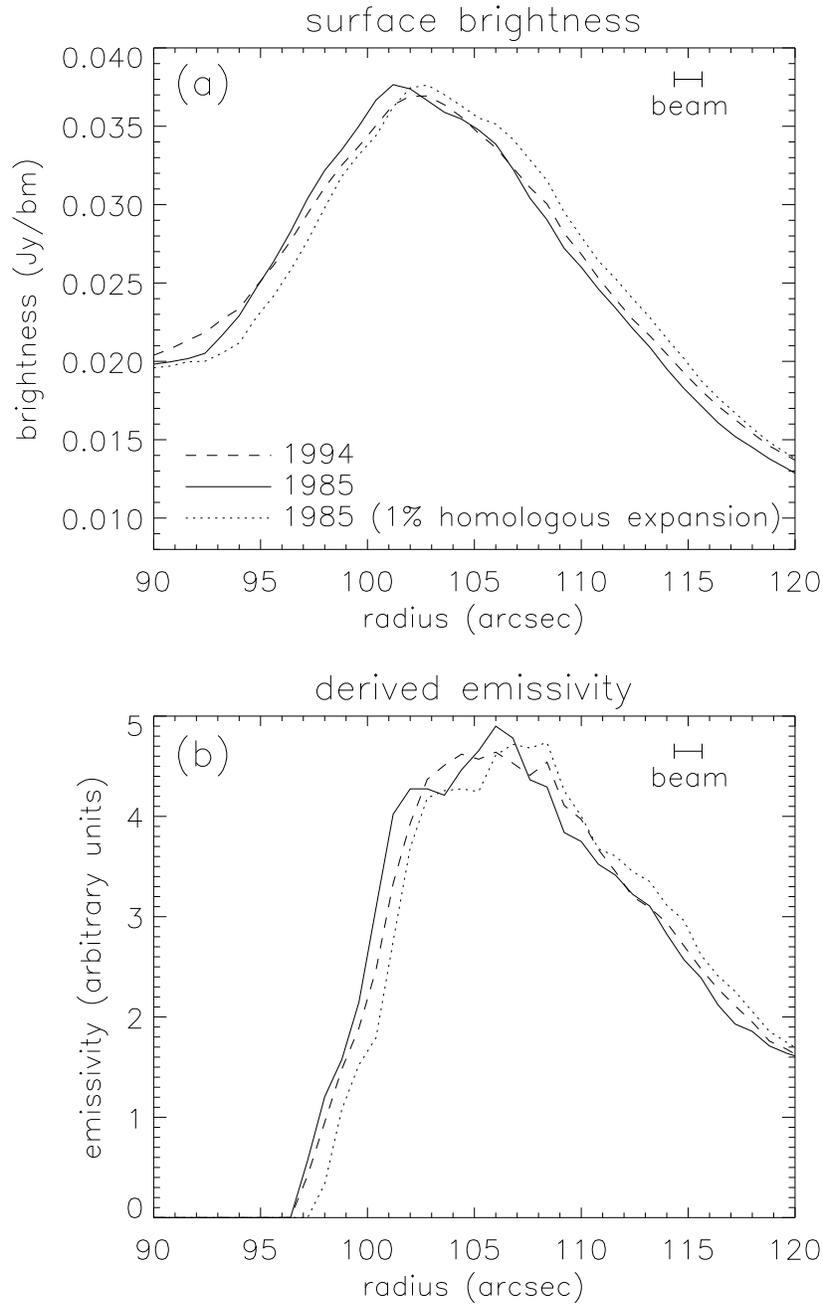}
\caption{(a): angle averaged radial brightness profiles of the radio bright 
ring for the 1985 and 1994 data.  Also shown is the profile of the 1985 
data with a 1\% homologous expansion simulating the 0.11\% yr$^{-1}$ 
expansion rate measured for the bright ring \citep{krg98} over 9.2 years.  
(b): derived emissivity profiles for the data in the top figure.  The beam 
FWHM is 1$\farcs$3.  
\label{profiles}}
\end{figure}

To calculate the emissivity profile, we deproject the observed line-of-sight 
integrated radio brightness as outlined in \citet{gkr01}.  The general method 
is to assume that the emissivity (at least for a limited range in azimuth, in 
this case) can be modeled as a set of thin, uniform shells.  We perform an 
iterative decomposition of the brightness profile as a function of radius 
into 1 pixel (0$\farcs$8) wide constant emissivity shells.  The iteration 
proceeds until the residual brightness is lowered to 1\% of its original peak 
value.  We then sum all of the emissivity components at each radius to 
generate an average emissivity for the shell at that radius.  For this 
analysis, we chose the azimuth range from 115$\degr$ to 250$\degr$, using a 
center of $23^{\mathrm{h}}23^{\mathrm{m}}25\fs44$ and 
$+58\degr48\arcmin52\farcs3$ (J2000) that best fits the radius of curvature 
of the radio bright ring.  Over this azimuth range, the radio bright ring is 
rather uniform.  Figure \ref{profiles} shows the angle-averaged radial 
brightness profiles and the corresponding emissivity profiles.  Also shown is 
the profile of the 1985 data with a 1\% homologous expansion simulating the 
0.11\% yr$^{-1}$ expansion rate measured for the bright ring \citep{krg98} 
over 9.2 years.

We measured the motion of the brightness profiles by minimizing $\chi^2$ of 
the difference between the profiles at each epoch as a function of radial 
shift and amplitude scaling factor.  For the emissivity profiles, we split the 
azimuth range into five segments and then minimized the variance of the 
differences between the profiles at each epoch as a function of radial 
shift and amplitude scaling factor for each segment.  We only used the inside 
edge and peak of the emissivity profiles because the position of the sharp 
rise could be measured accurately.  

The new measurements of the angle-averaged brightness and emissivity profiles 
of the radio bright ring show expansion rates of 0.02 $\pm$ 0.03\% yr$^{-1}$ 
and 0.07 $\pm$ 0.03\% yr$^{-1}$, respectively.  We consider the emissivity 
profile measurements to be more reliable than the brightness profiles since 
they proved to be robust to changes in amplitude scale.  Also, the more rapid 
fading of the bright ring, compared to the plateau, will bias the proper 
motion measurements of the brightness profile.

\section{Expansion Measurement Comparisons}

Table \ref{tbl3} is a summary of the most recent X-ray, optical, and radio 
expansion rate measurements.  For completeness, expansion parameters, 
expansion timescales, and velocities (using a distance of 3.4 kpc) are also 
shown along with average Doppler velocities where appropriate.

Although ours is the first direct measurement of the forward shock expansion, 
inferred forward shock velocities have been reported.  \citet{vbk98} assumed 
(correctly) that the X-ray bright ring expansion rate is the same as the 
forward shock expansion rate and derived a velocity of 5200 km s$^{-1}$.  
\citet{wbv02} use X-ray Doppler measurements to infer a forward shock velocity 
of 4000 km s$^{-1}$ -- 20\% less than our measurement for the forward shock 
fragments.

There have been many expansion measurements of the bright ring, with most 
sensitive to small-scale features.  The expansion rates for the different 
wavebands are in the approximate ratio of 3:2:1 for the optical, X-ray, and 
radio respectively (Thorstensen, Fesen, \& van den Bergh 2001 and references 
therein) indicating various degrees of deceleration for the material emitting 
in those wavebands.  Our new expansion measurement for the radio bright ring 
using the large-scale, diffuse emission is slower than the other measures of 
the bright ring.  The radio bright ring small-scale expansion measurements 
with azimuthal variations are indicated in Figure \ref{expplot} as well as 
the median value of the X-ray bright ring small-scale expansion measurements.

The bulk expansion of Cas A has been measured in the radio using the Bessel 
function nulls in the visibility plane (which is the Fourier transform of the 
sky brightness distribution).  The reported expansions of 
$\approx$0.11\% yr$^{-1}$ \citep{ar95} and $\approx$0.22\% yr$^{-1}$ 
\citep{ag99} are higher than our new angle-averaged emissivity expansion rate 
(0.07\% yr$^{-1}$).  The cause of this discrepancy is likely due to the more 
rapid fading of the bright ring compared to the plateau.  The positions of 
the nulls are affected by the differential change in brightness causing a 
bias towards higher expansion rates.  A direct comparison with our 
measurements is therefore not possible at this stage.

\section{Discussion}
\label{sec:dis}

For the purpose of this discussion, we will assume that the outer, thin 
X-ray continuum fragments represent the forward shock \citep{gkr01}.  
Then, in an effort to understand the dynamical picture presented by the 
expansion measurements of the X-ray and radio bright ring small-scale 
features and the angle-averaged radio bright ring emissivity profile, we 
use a dynamical simulation by \citet{tm99} that follows the forward and 
reverse shock evolution of nonradiative SNRs.  Their simulations trace SNR 
evolution from the self-similar ejecta-dominated stage to the self-similar 
Sedov-Taylor stage with a period of non--self-similar behavior in between.  
They concentrate on spherically symmetric ejecta profiles described by 
power-law density profiles expanding into a uniform ambient medium. 

\begin{figure}
\epsscale{0.55}
\plotone{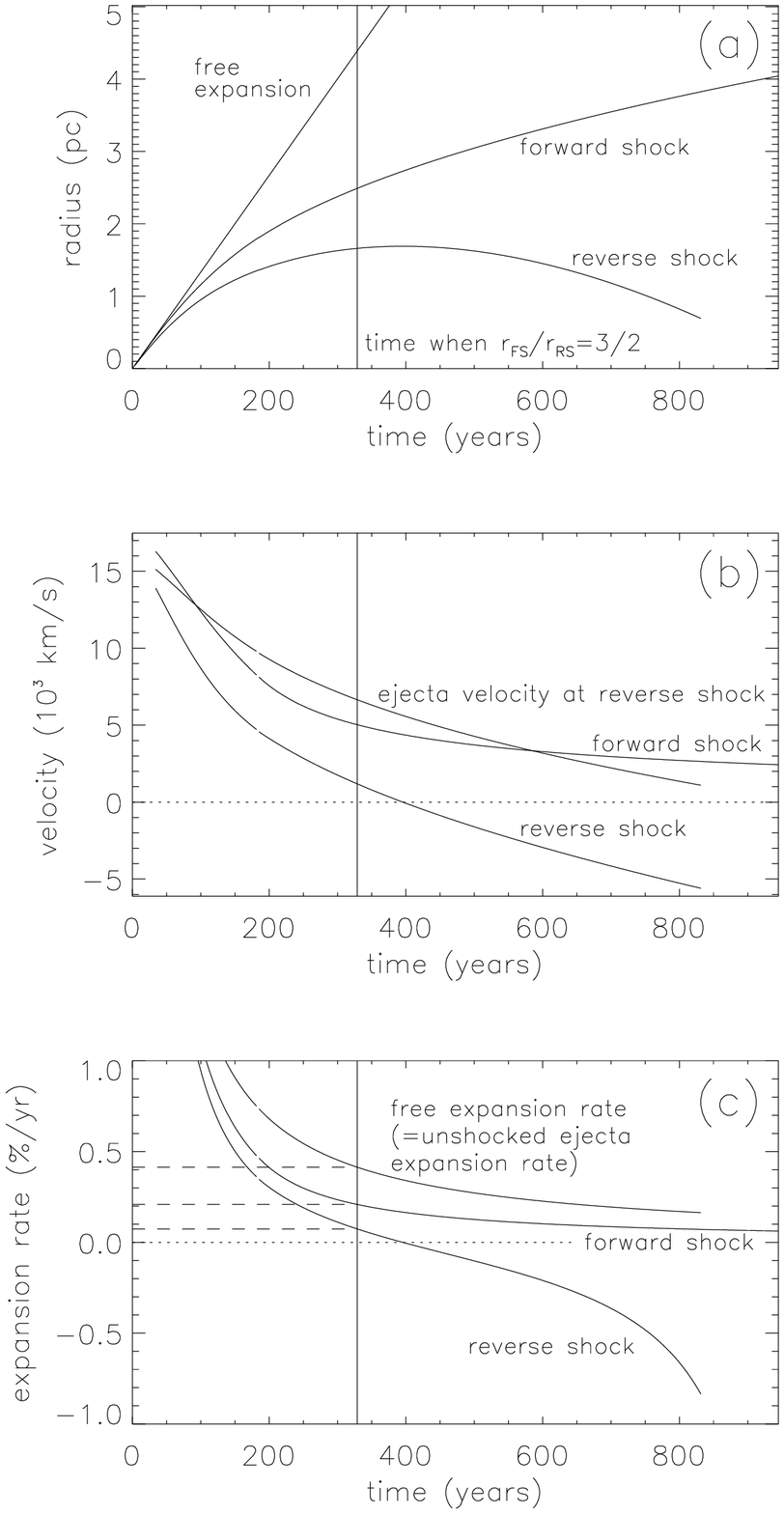}
\caption{Dynamical simulation from \citet{tm99} for a spherically 
symmetric nonradiative SNR with a uniform ejecta density profile expanding 
into a uniform ambient medium.  The time axis is normalized so that 
the time when $r_{FS}/r_{RS}=3/2$ (the current observed ratio \citep{gkr01}) 
is the age of the remnant as determined by \citet{tfv01}.  The radius, 
velocity, and expansion rate axes are normalized to the median values of the 
forward shock fragments plotted in Figure \ref{expplot}. \label{tm}}
\end{figure}

For an initial first order comparison, we use their analytical solutions for 
evolution in a uniform ambient medium with a uniform ejecta profile (both 
described by a density power-law index ($n$ for ejecta, $s$ for ambient 
medium) of 0) which is plotted in Figure \ref{tm}.  Hereafter, this 
simulation will be referred to as T\&M(0,0).  Although a higher ejecta 
density profile ($n\ge7$) is expected for core collapse SNe, the solutions 
at $n=7$ are very similar to the $n=0$ solutions during the Sedov-Taylor 
stage and thus do not change our conclusions.  Although Truelove \& McKee 
use dimensionless units, we have normalized the time axis in Figure 
\ref{tm} so that the time when $r_{FS}/r_{RS}=3/2$ (the current observed 
ratio \citep{gkr01}) is the age of the remnant as determined by 
\citet{tfv01} using outlying optical knots.  The radius, velocity, and 
expansion rate axes in Figure \ref{tm} are normalized to the median values 
of the forward shock fragments.  Following this normalization, the predicted 
T\&M(0,0) values for $r_{RS}$, $v_{RS}$, and reverse shock expansion rate 
are 1.66 pc, 1187 km s$^{-1}$, and 0.074\% yr$^{-1}$, respectively.  The 
T\&M(0,0) predicted reverse shock expansion rate is indicated on Figure 
\ref{expplot}.
 
We can convert the T\&M(0,0) simulation from dimensionless to physical 
units by assuming that the energy of the explosion is 10$^{51}$ erg and the 
ambient density is 3.2 cm$^{-3}$ (derived by Truelove \& McKee from 
\citet{bsb96}).  The current age of 331 years, as defined by outlying 
optical knots \citep{tfv01}, then results in an ejecta mass of 
1.4 M$_{\sun}$, a forward shock radius of 2.34 pc, a forward shock velocity 
of 3595 km s$^{-1}$, a reverse shock radius of 1.56 pc, a reverse shock 
velocity of 1270 km s$^{-1}$, and a swept up mass of 5.9 M$_{\sun}$.  The 
SNR is in the Sedov-Taylor stage and the reverse shock has penetrated into 
the inner, uniform core.  The low ejecta mass is consistent with a massive 
star that has lost most of its mass through winds.

The similarity between our new angle-averaged radio bright ring expansion 
rate and the predicted reverse shock expansion rate of the T\&M(0,0) 
simulation supports our identification of the reverse shock motion in Cas A.  
In this picture, the sharp rise of the radio emissivity profile indicates the 
location of the reverse shock.  The reverse shock being at the location of 
the bright ring is not unprecedented.  \emph{Hubble Space Telescope} imaging 
of Cas A shows finger-like structures in the bright ring that are thought to 
be Rayleigh-Taylor instabilities created in the interface between the reverse 
shock and the clumped ejecta \citep{fmc01}.  The motion of the emissivity 
profile would still represent the motion of the reverse shock even if there 
is some characteristic delay between the passage of the reverse shock and the 
onset of the radio emission.  The previous measurements of the X-ray and 
radio bright ring failed to measure the reverse shock motion because they 
were dominated by the behavior of the small-scale knots and filaments (B. 
Koralesky private communication).

The similar expansion rates of the X-ray bright ring and X-ray forward shock 
fragments can be explained if the small-scale bright ring features represent 
ejecta in various stages of deceleration after \emph{passing through} the 
reverse shock.  Some ejecta may be moving faster and some ejecta may be moving 
slower than the forward shock.  Note on Figure \ref{tm}c that the fastest 
moving ejecta are in free expansion and thus always have a larger 
\emph{expansion rate} than the forward shock fragments in Figure 
\ref{expplot} regardless of their velocity.  We also are in a regime where 
the fastest ejecta also have higher \emph{velocities} than the forward shock 
(Figure \ref{tm}b).  Indeed, dense clumps of optical ejecta are observed 
outside of the forward shock \citep{fes01}.  Presumably, these clumps are so 
dense and present such a small cross-section that they are not appreciably 
decelerated whereas the forward shock does experience deceleration by 
interaction with the circumstellar medium (CSM) \citep{fes01}.

The above explanation can be extended to the optical and radio small-scale 
bright rings by considering the physical properties of the emitting material.  
The optical fast moving knots are dense clumps of ejecta, and have experienced 
little deceleration by the reverse shock before they become radiative 
\citep{rhf95}.  The separation of the optical, X-ray, and radio emitting 
material by dynamical state has also been suggested by \citet{ajr94} and 
\citet{hsp01}.  The radio emission arises from synchrotron radiation which 
requires amplification of magnetic fields to be strong enough for the 
emission to be seen.  This amplification results from the turbulence 
associated with ejecta deceleration.  If significant deceleration is needed 
for amplification, the small-scale features in the radio would be 
decelerated the most.  \citet{wbv02} use a similar argument for dense 
``bullets'' of ejecta material that have penetrated the forward shock and 
have formed bow shocks in the ambient CSM.  The optical emission from these 
bullets arises from shocks.  The X-ray emission comes from material that has 
ablated off of the bullets, been heated by the bow shock, and drifted into 
the wake of the bullet.  The radio emission then results when the bullets 
decelerate and electrons accelerated by the bow shock radiate in 
magnetic fields amplified by shearing between the bullets and the CSM.

\begin{figure}
\epsscale{0.8}
\plotone{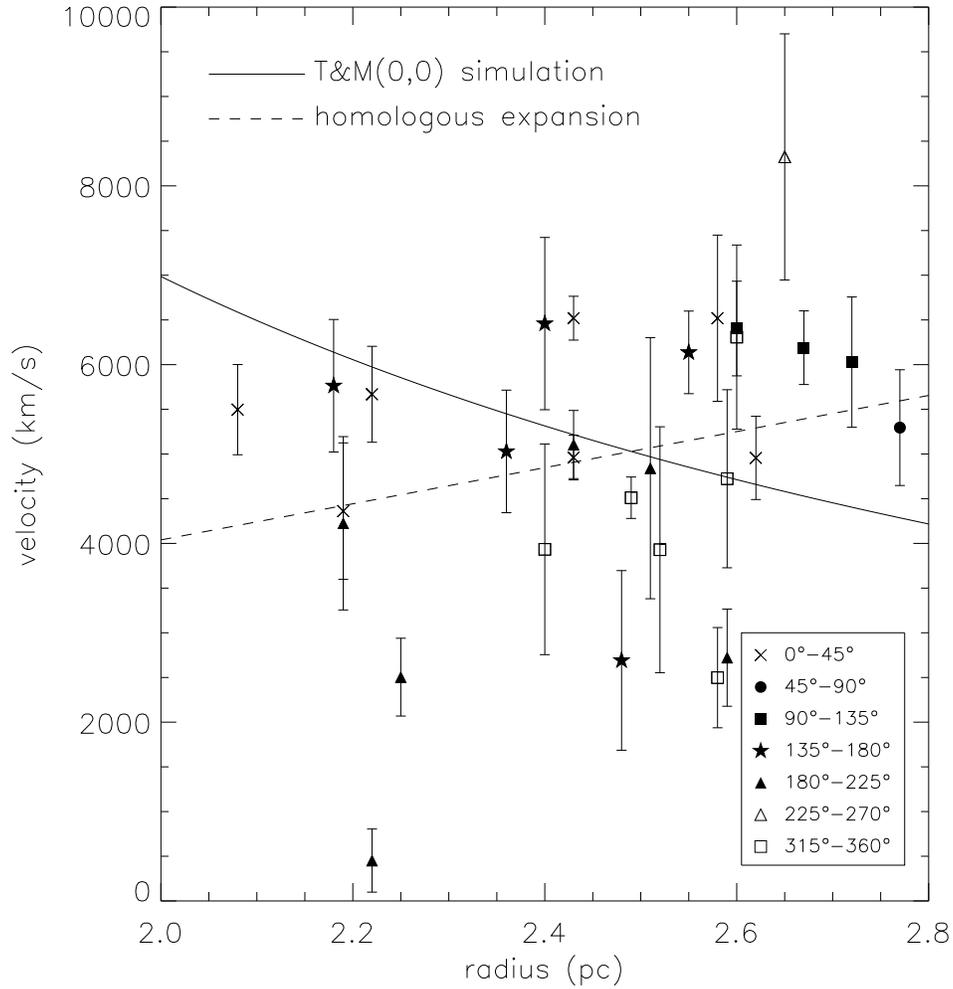}
\caption{Velocity vs. radius for the forward shock fragments.  Solid line:  
T\&M(0,0) simulation.  Dashed line:  homologous expansion normalized to the 
median velocity and radius of the forward shock fragments.  The symbols 
represent 45$\degr$ azimuth sectors.  No measurements were taken 
between 270$\degr$ and 315$\degr$.  \label{rv}}
\end{figure}

Although we find that the T\&M(0,0) simulation is helpful for understanding 
the average behavior of Cas A, there are important areas where the 
simulation and the observations disagree.  One is the free expansion rate.  
Figure \ref{tm}c shows that the T\&M(0,0) free expansion rate is 
0.41\% yr$^{-1}$ -- 36\% higher than the observed free expansion rate 
of 0.3\% yr$^{-1}$ \citep{tfv01}.  Another example of the limitations of the 
T\&M(0,0) simulation is in describing the differences among the expansion 
rates of the forward shock fragments.  In principle, the evolutionary state 
of any part of the SNR is determined from the local ratio of swept up mass 
to ejecta mass.  Those regions of higher CSM density should produce more 
deceleration than those regions of lower CSM density, so the radius and 
velocity of the individual forward shock fragments should be at different 
stages on the evolutionary path.  We plot in Figure \ref{rv} the radius and 
velocity measurements of the forward shock fragments, labelled by 45$\degr$ 
azimuth sectors, along with the velocity evolution of the T\&M(0,0) 
simulation.  The individual forward shock fragments do not follow the 
T\&M(0,0) evolutionary trend.

\begin{figure}
\epsscale{0.8}
\plotone{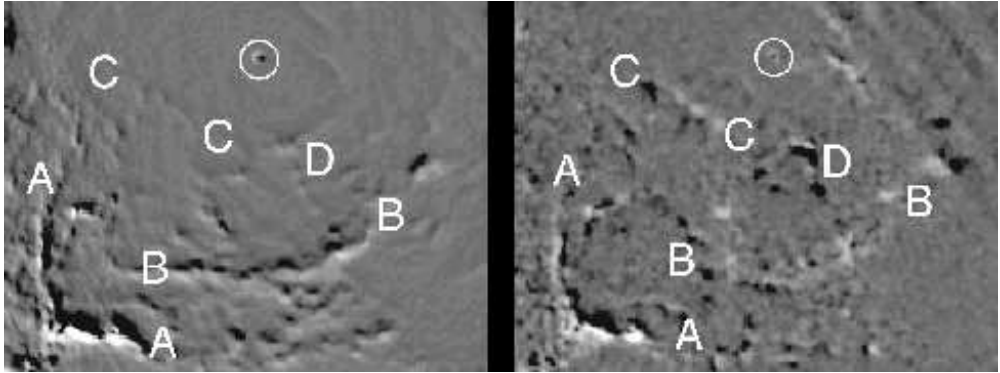}
\caption{Left: difference image between the first epoch \emph{Chandra} 
X-ray image and the first epoch image expanded by 
0.2\%.  Right: difference image between the second and first epoch 
\emph{Chandra} X-ray images.  The location of the point source is circled.  
Bright indicates direction of motion.  The stripes in the upper right corner 
are due to the node boundaries and dead columns on the ACIS-S3 chip.  
\label{expdiff}}
\end{figure}

We also consider a homologous expansion model for the forward shock 
variations.  This model has zero velocity at zero radius and the slope is 
determined by the median forward shock fragment velocity at the median 
radius.  Homologous expansion also fails to describe variations in the 
forward shock fragments as shown in Figure \ref{rv}.  Homologous expansion,  
or any other monotonic relationship between velocity and radius, also cannot 
be used to describe the motions of the interior X-ray knots and filaments.  
The complicated interior motions are shown in Figure \ref{expdiff} where we 
compare a 0.2\% homologous expansion (left) with the actual expansion 
(right).  For reference, the point source is in the center of the circle and 
bright indicates the direction of motion.  Both filament C and knot D are 
moving inwards while filament A is moving outwards.  Filament B is 
complicated, but parts of it are also moving inward.  The motions shown here 
and in Figure \ref{casdiff} cannot be accounted for with a different choice 
of registration between the two X-ray epochs.

We have also briefly considered two models with more realistic CSM density 
profiles.  The first model is that of \citet{bsb96} in which the 
bright ring represents part of the red supergiant wind compressed by the 
blue supergiant wind from the progenitor.  In this model, the bulk of the 
X-ray emission in the bright ring comes from the swept up CSM shell.  
This model predicts that the velocity of the bright ring is slightly 
faster than the forward shock, which is not observed.  Also, the predicted 
forward shock radius is only 1.07 times the radius of the bright ring compared 
to the actual value of 1.5 \citep{gkr01}.  This model is also limited in that 
it does not address the differences in proper motion between the X-ray, 
radio, and optical features.  The second model is an extension of the models 
of \citet{tm99} to an $n$=9 ejecta density profile with an $s$=2 CSM density 
profile (M. Laming private communication).  This model predicts the correct 
ratio between the free expansion and forward shock expansion rates.  However, 
the predicted reverse shock expansion rate is twice that of our measured 
value.  Both of these models predict that the bright ring is closer to the 
forward shock than is currently measured and both predict that the reverse 
shock is interior to the bright ring.  Future hydrodynamic models with 
realistic CSM and ejecta density profiles should include predictions for 
the dynamics of all of the emitting plasmas.

There are numerous indications of an asymmetric SNe and/or asymmetric 
expansion.  \citet{fes01} found high-velocity, sulfur-rich optical ejecta 
distributed along the ``jet'' axis to the northeast and southwest of the 
remnant beyond the forward shock radius.  \citet{wbv03} find that most of 
the X-ray emitting material is concentrated within a double cone that is 
oriented at $-55\degr$ from north and 50$\degr$ out of the plane of 
the sky with the northern material further away from the observer than the 
southern material.  \citet{rl03} require an asymmetric explosion to explain 
the observed $^{44}$Ti flux.  One would expect explosion or expansion 
asymmetries to be evident in proper motion measurements of the bright ring 
and forward shock.  Indeed, there are significant variations in the X-ray 
and radio expansion rates of the bright ring ($\sim$50\%) as a function of 
azimuth \citep{krg98,vbk98}, however they do not show the clear bipolarity 
seen by \citet{fes01} and \citet{wbv03}.  There are also significant 
variations in the expansion rates of the forward shock fragments as shown in 
Figure \ref{expplot}; however we do not have adequate azimuthal sampling 
to say at this time if the expansion rates along the ``jet'' axis are 
significantly different from the rest of the SNR.

\section{Conclusions \& Future Work}
\label{sec:con}

We have measured radial proper motions for the X-ray forward shock fragments 
in Cas A.  We have also re-analyzed measurements of the radio bright ring 
using angle-averaged emissivity profiles to isolate and measure the motion 
of the reverse shock.  Our conclusions are as follows:

\begin{enumerate}

\item The thin, continuum dominated X-ray filaments associated with the 
forward shock show expansion rates from 0.02\% yr$^{-1}$ to 
0.33\% yr$^{-1}$.  Most of the X-ray forward shock fragments have been 
significantly decelerated from the free expansion rate of 0.3\% yr$^{-1}$ 
\citep{tfv01}.  

\item The median expansion rate of the forward shock fragments is 
0.21\% yr$^{-1}$ which is the same as the expansion rate measured for the 
small-scale X-ray features in the bright ring, twice the expansion rate 
of the small-scale radio features of the bright ring \citep{krg98,vbk98}, 
and 2/3 the expansion rate of the small-scale optical features of the bright 
ring \citep{tfv01}.

\item The reverse shock expansion rate, measured using angle-averaged radio 
emissivity profiles of the bright ring, is 0.07\% yr$^{-1}$ -- slower than 
the expansion rate of the small-scale radio features of the bright ring.

\item There is general agreement between the global relations of the forward 
and reverse shocks (median velocity ratios and median radius ratios) and the 
hydrodynamical simulations of \citet{tm99} who model expansion of uniform 
ejecta into a uniform ambient medium (T\&M(0,0)).  

\item The interpretation of the small-scale bright ring features in the 
optical, X-ray, and radio as ejecta in various stages of deceleration after 
passage through the reverse shock naturally establishes a velocity gradient 
between the three wavebands from the densest, fastest optical knots to the 
most decelerated radio features.

\item The T\&M(0,0) simulation does not adequately describe the observed 
relationship between free expansion rate and the forward shock median 
expansion rate.  Nor does the simulation describe the observed relation 
between the radius and velocity of forward shock fragments at different 
azimuths.

\item The motions of the X-ray material at the forward shock, in the bright 
ring, and in the interior of the remnant are complex -- the interior 
involving inward as well as outward motions -- and cannot be modeled with a 
homologous expansion.

\item Because there was considerable mass loss from Cas A's progenitor, models 
that include more realistic CSM density profiles such as that of \citet{bsb96} 
are important for understanding the dynamics of the SNR.

\end{enumerate}

\acknowledgments

This work was supported at the University of Minnesota under NSF grant AST 
96-19438 and Smithsonian grant SMITHSONIAN/GO1-2051A/NASA.  We thank Una 
Hwang, Rob Petre, and especially Paul Plucinsky for considerable help 
setting up the second epoch \emph{Chandra} X-ray observations.  We are 
grateful to Martin Laming for providing his $s$=2 solutions and for very 
helpful discussions.  We appreciate discussions with Maxim Markevitch about 
the point spread function and plate scale of the ACIS instrument.  We also 
thank Aneta Siemiginowska and Elizabeth Galle at the Chandra X-ray Center for 
help processing (and reprocessing) both epoch X-ray observations.

\clearpage

\begin{deluxetable}{lll}
\tabletypesize{\scriptsize}
\tablecaption{\emph{Chandra} ACIS-S3 Observational Parameters \label{tbl1}}
\tablewidth{0pt}
\tablehead{
\colhead{Parameter} & 
\colhead{2000 Value} & 
\colhead{2002 Value}}
\startdata
Total Observation Time (ks) & 52.4 & 51.4 \\
Target RA (deg) & 350.861250 & 350.861250 \\
Target Dec (deg) & 58.817500 & 58.817500 \\
Pointing RA (deg) & 350.917812 & 350.917312 \\
Pointing Dec (deg) & 58.792819 & 58.793648 \\
Pointing Roll (deg) & 323.387104 & 323.391738 \\ 
Chip Temperature (K) & 153.123627 & 153.284805 \\
Data Mode & Graded & Graded \\
Read Mode & Timed & Timed 
\enddata
\end{deluxetable}

\clearpage

\begin{deluxetable}{cccccccc}
\tabletypesize{\scriptsize}
\tablecaption{Forward Shock Fragment Measurements \label{tbl2}}
\tablewidth{0pt}
\tablehead{
\colhead{Region} & 
\colhead{Azimuth} & 
\colhead{Radius} &
\colhead{Expansion Rate} &
\colhead{Expansion Rate Error} &
\colhead{Radius\tablenotemark{a}} &
\colhead{Velocity\tablenotemark{a}} &
\colhead{Velocity Error\tablenotemark{a}}
\\
\colhead{} &
\colhead{(degrees)} &
\colhead{(arcsec)} &
\colhead{(\% yr$^{-1}$)} &
\colhead{(\% yr$^{-1}$)} &
\colhead{(pc)} &
\colhead{(km s$^{-1}$)} &
\colhead{(km s$^{-1}$)}}
\startdata
\phn1 & \phn51.8 & 167.83 & 0.196 & 0.024 & 2.77 & 5296 & \phn647 \\
\phn2 & \phn38.8 & 147.59 & 0.274 & 0.010 & 2.43 & 6520 & \phn245 \\
\phn3 & \phn42.0 & 132.75 & 0.204 & 0.036 & 2.19 & 4361 & \phn763 \\
\phn4 & \phn41.1 & 126.42 & 0.270 & 0.025 & 2.08 & 5497 & \phn505 \\
\phn5 & \phn37.0 & 134.91 & 0.261 & 0.025 & 2.22 & 5668 & \phn537 \\
\phn6 & \phn20.8 & 156.34 & 0.259 & 0.037 & 2.58 & 6519 & \phn930 \\
\phn7 & \phn12.2 & 158.80 & 0.194 & 0.018 & 2.62 & 4956 & \phn465 \\
\phn8 & \phn\phn1.8 & 147.42 & 0.209 & 0.011 & 2.43 & 4961 & \phn250 \\
\phn9 & 353.0 & 156.65 & 0.099 & 0.022 & 2.58 & 2498 & \phn561 \\
10 & 347.3 & 157.17 & 0.186 & 0.039 & 2.59 & 4723 & \phn996 \\
11 & 347.0 & 153.00 & 0.159 & 0.056 & 2.52 & 3929 & 1376 \\
12 & 348.1 & 145.80 & 0.167 & 0.050 & 2.40 & 3934 & 1179 \\
13 & 341.2 & 157.45 & 0.249 & 0.041 & 2.60 & 6308 & 1030 \\
14 & 338.4 & 151.01 & 0.185 & 0.010 & 2.49 & 4511 & \phn232 \\
15 & 234.3 & 160.77 & 0.321 & 0.053 & 2.65 & 8324 & 1377 \\
16 & 208.2 & 134.96 & 0.021 & 0.016 & 2.22 & \phn452 & \phn353 \\
17 & 192.3 & 157.12 & 0.107 & 0.022 & 2.59 & 2722 & \phn542 \\
18 & 188.9 & 152.39 & 0.197 & 0.060 & 2.51 & 4841 & 1461 \\
19 & 186.7 & 147.30 & 0.215 & 0.016 & 2.43 & 5106 & \phn382 \\
20 & 191.4 & 136.53 & 0.114 & 0.020 & 2.25 & 2504 & \phn435 \\
21 & 180.4 & 132.84 & 0.197 & 0.045 & 2.19 & 4225 & \phn970 \\
22 & 175.7 & 132.22 & 0.270 & 0.035 & 2.18 & 5764 & \phn741 \\
23 & 154.2 & 150.29 & 0.111 & 0.042 & 2.48 & 2691 & 1005 \\
24 & 152.5 & 145.90 & 0.275 & 0.041 & 2.40 & 6459 & \phn964 \\
25 & 144.7 & 142.93 & 0.218 & 0.030 & 2.36 & 5029 & \phn686 \\
26 & 136.9 & 154.85 & 0.246 & 0.019 & 2.55 & 6137 & \phn462 \\
27 & 125.0 & 157.73 & 0.252 & 0.021 & 2.60 & 6404 & \phn529 \\
28 & 122.8 & 165.01 & 0.227 & 0.027 & 2.72 & 6028 & \phn729 \\
29 & 108.5 & 162.07 & 0.237 & 0.016 & 2.67 & 6190 & \phn412 \\
mean & & 149.18 & 0.204 & & 2.46 & 4916 & \\	
median & & 151.01 & 0.209 & & 2.49 & 5029 & \\
rms & & \phn11.03 & 0.066 & & 0.18 & 1623 & \\
error in mean & & \phn\phn2.05 & 0.012 & & 0.03 & \phn301	
\enddata
\tablenotetext{a}{Assuming distance of 3.4 kpc}
\end{deluxetable}

\clearpage

\begin{deluxetable}{lccccl}
\tabletypesize{\scriptsize}
\tablecaption{Expansion Measurement Comparisons \label{tbl3}}
\tablewidth{0pt}
\tablehead{
\colhead{} &
\colhead{Expansion} & 
\colhead{} & 
\colhead{Expansion} &
\colhead{Expansion} &
\colhead{}
\\
\colhead{Measurement\tablenotemark{a}} &
\colhead{Rate} &
\colhead{Velocity} &
\colhead{Parameter} &
\colhead{Timescale} &
\colhead{Reference}
\\
\colhead{} &
\colhead{(\% yr$^{-1}$)} &
\colhead{(km s$^{-1}$)} &
\colhead{$m$} &
\colhead{(years)} &
\colhead{}}
\startdata
Forward Shock & 0.21 & 5029 & 0.69 & \phn476 & this work \\
\\
Bright Ring: & & & & & \\
\phm{B}Optical small scale (Doppler) & \nodata & 5290 & \nodata & \nodata & \citet{rhf95} \\
\phm{B}Optical small scale & 0.30 & 4842 & 0.99 & \phn338 & \citet{tfv01} \\
\phm{B}X-ray small scale (Doppler) & \nodata & 2500 & \nodata & \nodata & \citet{hsp01, wbv02} \\
\phm{B}X-ray small scale & 0.21 & 3500 & 0.66 & \phn476 & \citet{krg98, vbk98} \\
\phm{B}Radio small scale & 0.11 & 1750 & 0.35 & \phn909 & \citet{krg98} \\
\phm{B}Radio large scale & 0.07 & 1155 & 0.22 & 1429 & this work
\enddata
\tablenotetext{a}{All measurements by proper motion unless noted}
\end{deluxetable}

\clearpage

\clearpage

\clearpage

\clearpage

\clearpage

\clearpage

\clearpage

\begin{thebibliography}{}

\bibitem[Ag\"{u}eros \& Green(1999)]{ag99} Ag\"{u}eros, M. A. \& Green, 
D. A. 1999, \mnras, 305, 957 

\bibitem[Anderson \emph{et al.}(1994)]{ajr94} Anderson, M., Jones, T. W., 
Rudnick, L., Tregillis, I. L., \& Kang, H. 1994, \apj, 421, L31

\bibitem[Anderson \& Rudnick(1995)]{ar95} Anderson, M. \& Rudnick, L. 1995, 
\apj, 441, 307

\bibitem[Anderson \emph{et al.}(1991)]{arl91} Anderson, M., Rudnick, L., 
Leppik, P., Perley, R., \& Braun, R. 1991, \apj, 373, 146

\bibitem[Borkowski \emph{et al.}(1996)]{bsb96} Borkowski, K. J., Szymkowiak, 
A. E., Blondin, J. M., \& Sarazin, C. L. 1996, \apj, 466, 866

\bibitem[Fesen(2001)]{fes01} Fesen, R. A. 2001, \apjs, 133, 161

\bibitem[Fesen \emph{et al.}(2001)]{fmc01} Fesen, R. A., Morse, J. A., 
Chevalier, R. A., Borkowski, K. J., Gerardy, C. L., Lawrence, S. S., \& van 
den Bergh, S. 2001, \aj, 122, 2644

\bibitem[Gotthelf \emph{et al.}(2001)]{gkr01} Gotthelf, E., Koralesky, B., 
Rudnick, L., Jones, T., Hwang, U., \& Petre, R. 2001, \apj, 552, L39

\bibitem[Hughes \emph{et al.}(2000)]{hrb00} Hughes, J. P., Rakowski, C. E., 
Burrows, D. N., \& Slane, P. O. 2000, \apj, 528, L109

\bibitem[Hwang, Holt, \& Petre(2000)]{hhp00} Hwang, U., Holt, S., \& Petre, 
R. 2000, \apj, 537, L119

\bibitem[Hwang \emph{et al.}(2001)]{hsp01} Hwang, U., Szymkowiak, A. E., 
Petre, R., \& Holt, S. S. 2001, \apj, 560, L175

\bibitem[Koralesky \emph{et al.}(1998)]{krg98} Koralesky, B., Rudnick, L., 
Gotthelf, E., \& Keohane, J. 1998, \apj, 505, L27

\bibitem[Reed \emph{et al.}(1995)]{rhf95} Reed, J. E., Hester, J. J., Fabian, 
A. C., \& Winkler, P. F. 1995, \apj, 440, 706

\bibitem[Rothschild \& Lingenfelter(2003)]{rl03} Rothschild, R. E. \& 
Lingenfelter, R. E. 2003, \apj, 582, 257

\bibitem[Thorstensen, Fesen, \& van den Bergh(2001)]{tfv01} Thorstensen, J., 
Fesen, R., \& van den Bergh, S. 2001, \aj, 122, 297

\bibitem[Truelove \& McKee(1999)]{tm99} Truelove, J. \& McKee, C. 1999, \apjs, 
120, 299

\bibitem[Vink \emph{et al.}(1998)]{vbk98} Vink, J., Bloemen, H., Kaastra, 
J. S., \& Bleeker, J. A. M. 1998, \aap, 339, 201

\bibitem[Willingale \emph{et al.}(2003)]{wbv03} Willingale, R., Bleeker, 
J. A. M., van der Heyden, K. J., \& Kaastra, J. S. 2003, \aap, 
398, 1021

\bibitem[Willingale \emph{et al.}(2002)]{wbv02} Willingale, R., Bleeker, 
J. A. M., van der Heyden, K. J., Kaastra, J. S., \& Vink, J. 2002, \aap, 
381, 1039

\end{thebibliography}
\end{document}